\documentclass[12pt]{iopart}

\usepackage{bm,graphicx,iopams,amssymb,algorithm,algorithmic} 

\begin{document}
\title{Improved Neuronal Ensemble Inference with Generative Model and MCMC}

\author{Shun Kimura}
\address{Department of Mechanical Systems Engineering, Graduate School of Science and Engineering, Ibaraki University, Hitachi, Ibaraki  316-8511, Japan}

\author{Keisuke Ota}
 \address{Department of Neurochemistry, Graduate School of Medicine, The University of Tokyo, Bunkyo-ku, Tokyo 113-0033, Japan}
 \address{Laboratory for Haptic Perception and Cognitive Physiology, Center for Brain Science (CBS), RIKEN, Wako,
Saitama 351-0198, Japan}
 \ead{k-ota@m.u-tokyo.ac.jp}

\author{Koujin Takeda}
\address{Department of Mechanical Systems Engineering, Graduate School of Science and Engineering, Ibaraki University, Hitachi, Ibaraki  316-8511, Japan}
\ead{koujin.takeda.kt@vc.ibaraki.ac.jp}

\vspace{10pt}

\begin{abstract}
Neuronal ensemble inference is a significant problem in the study of biological neural networks. Various methods have been proposed for ensemble inference 
from experimental data of neuronal activity. Among them, Bayesian inference approach with generative model was proposed recently. However, this method requires large computational cost for appropriate inference. In this work, we give an improved Bayesian inference algorithm by modifying update rule in Markov chain Monte Carlo method and introducing the idea of simulated annealing for hyperparameter control. 
We compare the performance of ensemble inference between our algorithm and the original one, and discuss the advantage of our method.
\end{abstract}

%
%
%
%
%


\section{Introduction}
In recent study of biological neural networks, advanced recording technologies such as calcium imaging or high-performance electrode technology enable us to obtain neuronal activity data from thousands of neurons simultaneously
\cite{dana2019high, nguyen2019simultaneous, Jun2017}. 
Such activity data will reveal features of neural network, because neurons in the same neuronal ensemble tend to fire synchronously \cite{fries2005mechanism, lopes2011neuronal}.
In fact, there are some studies on the whole biological neural network structure 
using ensemble information \cite{friedrich2004multiplexing, palva2010neuronal, wang2019spike}.
Moreover, in neuroscience, an action of animal will be associated with a specific neuronal ensemble \cite{engelhardtcerebral}.
 Therefore, inference of neuronal ensembles is also significant for understanding action of animal.

Several conventional statistical methods have been applied to neuronal ensemble inference from activity data. 
For instance, one can identify ensembles by principal component analysis or singular value decomposition \cite{romano2015spontaneous, stringer2019high}. 
Their advantage is that they can effectively reduce dimension of large scale data.
However, prior knowledge on data is generally required for interpretation of result, and large computational cost is necessary.
As widely-used inference methods for ensembles, k-means clustering and spectral clustering are known. 
In k-means clustering, time series data of neuronal activity is mapped to the point in high-dimensional space. 
However, neuronal activity data is represented too sparsely in the space, which makes neuronal ensemble inference difficult.
Spectral clustering is an ensemble inference method for graph, where the connectivity among neurons is expressed as edges.
This method has also been applied to neuronal ensemble inference \cite{sakuma2016large}. 
However, the number of ensembles (=clusters) should be given in advance in this method, and dynamical behavior of neuronal activity is not taken into account.

One of the strategies to overcome above-mentioned problems
is Bayesian modeling. In recent work, Bayesian inference framework
with generative model of ensemble activity was proposed \cite{diana2019bayesian}, where
 neuronal ensembles are inferred from large-scale time series data of neuronal activity 
by Markov chain Monte Carlo (MCMC) method. 
In their method, the number of neuronal ensembles is not given in advance, 
but is inferred by Dirichlet process (DP) \cite{neal2000markov}.
However, this method still requires large computational cost due to appropriate choice of 
initial condition in MCMC. 
If one attempts to decrease computational cost by changing initial condition, this yields 
inappropriate result of Bayesian inference.

In the present work, we propose an improved algorithm for neuronal ensembles in order
to reduce computational cost and to avoid inappropriate result. 
First, we change the update rule in MCMC for controlling the number of ensembles. 
Second, we introduce the idea of simulated annealing for hyperparameter control. 
We check the performance of our method using synthetic neuronal activity data.
The result shows that our method can reproduce ground-truth ensembles correctly 
and work faster than the original.
We also apply our method to real activity data from mouse, whose result gives
appropriate biological neuronal ensembles.

\section{Theory}
\subsection{Bayesian inference model}
The framework of Bayesian inference is outlined here. Note that we basically follow the notation in the previous work 
\cite{diana2019bayesian}. 
In our model there are $N$ neurons, and each neuron has the label $i \in \{ 1, 2, \ldots, N\}$. 
The discrete time step is denoted by $k \in \{1, 2, \ldots, M\}$, and $M$ is the size of time frame. 
There are multiple neuronal ensembles in this model, and the label of neuronal ensemble
is denoted by $\mu \in \{1, 2, \ldots, A\}$, where $A$ is the total number of ensembles. 
The $i$th neuron belongs to one of the neuronal ensembles, which is expressed by the membership label 
$t_i \in \{1, 2, \ldots, A\}$. The $i$th neuron also has binary neuronal activity $s_{ik} \in \{0, 1\}$ at time $k$. 
Furthermore, neuronal ensemble has its "ensemble" activity:
the $\mu$th ensemble has binary ensemble activity $\omega_{k \mu} \in \{0, 1\}$ at time $k$. 
For $s_{ik}$ and $\omega_{k \mu}$, 
the value $1$ means active (firing) neuron/ensemble, while the value $0$ is inactive. 

The generative model for neuronal activity is given as the conditional joint probability, 
\begin{eqnarray}
 &         & P( \bm t, \bm \omega, \bm s|\bm n, \bm p, \bm \lambda) 
             \nonumber \\
 & \propto & \left(
             \prod_{i=1}^{N} n_{t_{i}}
             \right) 
             \cdot
             \left(
             \prod_{\mu=1}^{A} \prod_{k=1}^{M} 
             p_{\mu}^{\omega_{k \mu}}
             (1-p_{\mu})^{1-\omega_{k \mu}} 
             \right)
             \nonumber \\
 &  & \cdot
             \left(
             \prod_{i=1}^{N} \prod_{k=1}^{M} 
             [\lambda_{t_{i}}(\omega_{kt_{i}})]^{s_{ik}} 
             [1-\lambda_{t_{i}}(\omega_{kt_{i}})]^{1-s_{ik}}
             \right),
\label{generative}
\end{eqnarray}
where boldface letter represents the set of variables (e.g. $\bm t = \{t_1,t_2, \ldots, t_N\}$).
The meaning of generative model in equation (\ref{generative}) is as follows. 
First, neuronal membership label $t_i$ ($i \in \{1, \ldots, N\}$) is drawn 
from categorical distribution with probability $n_{\mu}$ for ensemble $\mu \in \{1, \ldots, A\}$.
Second, binary ensemble activity $\omega_{k \mu}$ is drawn independently 
from Bernoulli distribution with parameter $p_{\mu}$.
Third, binary neuronal activity $s_{ik}$ is also drawn from Bernoulli distribution, where
the parameter of Bernoulli distribution $\lambda_{t_i} (w_{k t_i})$ depends on the ensemble activity $w_{k t_i}$ 
of the corresponding ensemble $t_i$. 
The parameter $\lambda_{t_i} (w_{k t_i})$ can be regarded as 
the conditional probability for given ensemble activity $\omega_{kt_{i}}$, 
when we let $s_{ik} = 0\ {\rm or}\ 1$ in the third parenthesis on r.h.s. of equation (\ref{generative}),
\begin{eqnarray}
\lambda_{t_i}( \omega_{k t_i} ) &=& P(s_{ik}=1\ |\ \omega_{k t_i} ), \nonumber \\
1 - \lambda_{t_i}( \omega_{k t_i} ) &=& P(s_{ik}=0\ |\ \omega_{k t_i} ) \ \ {\rm for} \ \omega_{k t_i} \in \{ 0,1 \}.
\label{conductivity}
\end{eqnarray}
Namely, the parameters $\lambda_{t_i} (1)$ and $\lambda_{t_i} (0)$ represent 
the probabilities of neurons in the ensemble $t_i$ to be active
when the ensemble is active ($\omega_{k t_i}=1$) or inactive ($\omega_{k t_i}=0)$ at time $k$, respectively.
Hence, the parameter $\lambda_{t_i}$ describes coherence or incoherence (=noise) 
between neuronal activity $s_{ik}$ and ensemble activity $\omega_{k t_i}$.
  
\begin{figure}
\begin{center}
	\includegraphics[scale=0.70]{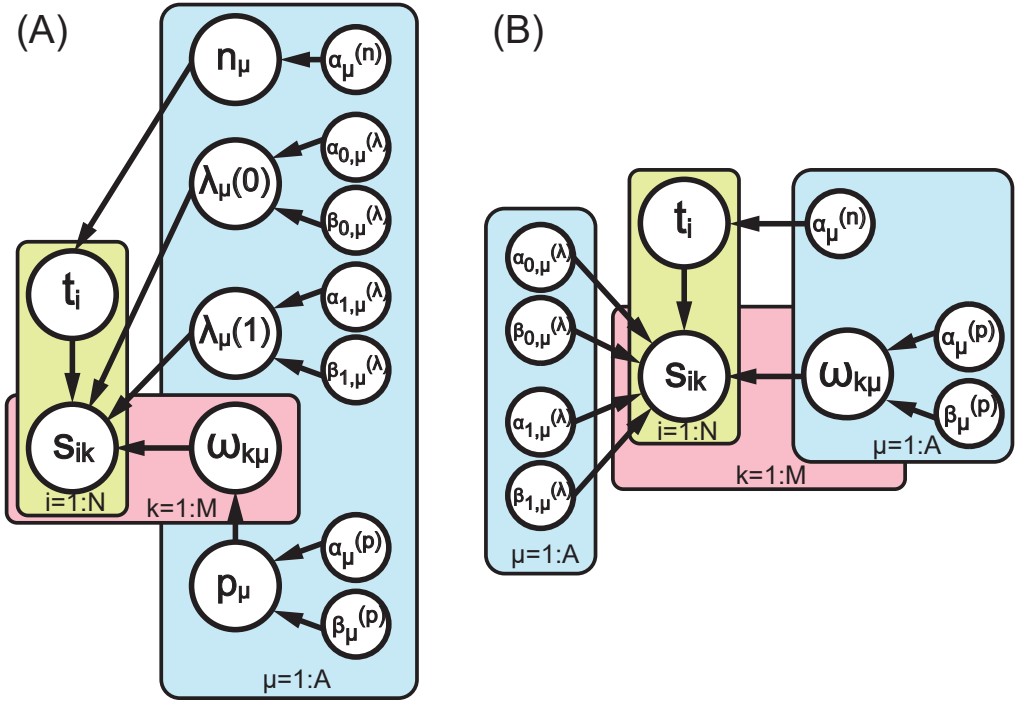}
	\caption{The relation among 
	         variables, parameters, and hyperparameters in Bayesian inference model:
	        (A) The full model. 
	        (B) The simplified model after integration out of 
	        $\bm n, \bm p, \bm \lambda$.}	 
\label{Fig.graphical}
\end{center}	                     
\end{figure}

 In addition, priors are also assumed for the model parameters $\bm p, \bm \lambda, \bm n$.
 For convenience of analysis, conjugate priors are chosen:
 the priors of ensemble activity rate $\bm p$ and conditional activity rate $\bm \lambda$ are chosen as 
 beta distribution (denoted by Beta), 
 while the prior of assigning probability $\bm n$ is Dirichlet distribution (by Dir),
\begin{eqnarray}
P (p_{\mu}) & = & {\rm Beta} \left( 
                             \alpha^{(p)}_{\mu}, \beta^{(p)}_{\mu} 
                             \right), \\
P (\lambda_{\mu}(z)) & = & {\rm Beta} \left( 
                                      \alpha_{z, \mu}^{(\lambda)},  
                                      \beta_{z, \mu}^{(\lambda)} 
                                      \right), \\
P(  n_{1}, \cdots, n_{A} ) & = & {\rm Dir} \left( 
                                           \alpha_{1}^{(n)},\cdots, 
                                           \alpha_{A}^{(n)} 
                                           \right),
\end{eqnarray}
where $\alpha^{(p)}_{\mu}, \beta^{(p)}_{\mu}, \alpha_{z,\mu}^{(\lambda)}, \beta_{z,\mu}^{(\lambda)}, \alpha_{\mu}^{(n)} $ ($z \in \{0,1\}$, $\mu \in \{1,2,\ldots, A\}$) are hyperparameters of priors. 
The relation among variables, parameters, and hyperparameters in our model is represented graphically in figure \ref{Fig.graphical}(A).

The model parameters can be integrated out $\{ \bm n, \bm p, \bm \lambda \}$ analytically. 
Integration over these parameters yields the joint probability as
\begin{eqnarray}
&  & P(\bm t, \bm \omega, \bm s) 
 =        \int d \bm n d \bm p d \bm \lambda\ 
            P(\bm t, \bm \omega, \bm s | \bm n, \bm p, \bm \lambda )
            P( \bm n, \bm p, \bm \lambda ) \nonumber \\
& \propto & \int d \bm n d \bm p d \bm \lambda
            \left(
            \prod_{i=1}^{N} n_{t_{i}}
            \right)
            \cdot
            {\rm Dir} \left( 
                      \alpha_{1}^{(n)}, \ldots, \alpha_{A}^{(n)}
			          \right)
		    \nonumber \\           
& &         \cdot 
            \left(
            \prod_{\mu=1}^{A} \prod_{k=1}^{M} 
            p_{\mu}^{\omega_{k\mu}}
            (1-p_{\mu})^{1-\omega_{k\mu}} 
            \right)
            \cdot 
            \left( 
            \prod_{\mu=1}^{A} {\rm Beta} 
            \left( 
            \alpha^{(p)}_{\mu}, \beta^{(p)}_{\mu}
			\right) 
			\right)
            \nonumber \\
& &         \cdot 
            \left( 
            \prod_{i=1}^{N} \prod_{k=1}^{M} 
            [\lambda_{t_{i}}(\omega_{k t_{i}})]^{s_{ik}} 
            [1-\lambda_{t_{i}}(\omega_{kt_{i}})]^{1-s_{ik}}
            \right) 
            \nonumber\\
& &         \hspace{4cm} \cdot 
            \left( 
            \prod_{\mu=1}^{A} \prod_{z \in \{ 0,1 \}}
            {\rm Beta} \left( 
                       \alpha_{z, \mu}^{(\lambda)},
			           \beta_{z, \mu}^{(\lambda)}
			           \right) 
	        \right)	\nonumber \\		                
& = &       \left(
            \frac{\mathcal{B}(\alpha_1^{(n)}+G_1,\ 
                  \alpha_2^{(n)}+G_2, \ldots,\ \alpha_A^{(n)}+G_A)}
                 {\mathcal{B}(\alpha_1^{(n)}, \alpha_2^{(n)}, 
                  \ldots, \alpha_A^{(n)})}
            \right) \nonumber \\ 
& &         \cdot \prod_{\mu=1}^{A}
            \left\{
            \frac{B(H_{\mu}, \bar{H}_{\mu})}
                 {B(\alpha_{\mu}^{(p)}, \beta_{\mu}^{(p)})}
            \prod_{z \in \{ 0, 1 \}}
            \frac{B(T_{\mu}^{z1}, T_{\mu}^{z0})}
                 {B(\alpha_{z,\mu}^{(\lambda)}, \beta_{z,\mu}^{(\lambda)})}     
            \right\},
\label{P(t, omega, s)}
\end{eqnarray}
where $B(\cdot, \cdot)$ is beta function and $\mathcal{B}$ is multivariate beta function defined by
\begin{equation}
\mathcal{B} ( x_{1}, \cdots, x_{A} )
= \frac{\prod_{k=1}^A {\Gamma}(x_{k})}
            {{\Gamma}(\sum_{k=1}^A x_{k})}.
\end{equation}
Several variables for the $\mu$th ensemble are also introduced in equation (\ref{P(t, omega, s)}), 
\begin{eqnarray}
G_{\mu} & = & \sum_{i=1}^{N} \delta_{\mu, t_{i}},
\quad \nonumber \\
H_{\mu} & = & \alpha^{(p)}_{\mu} + \sum_{k=1}^{M}\omega_{k\mu},
\quad \nonumber \\
\bar{H}_{\mu} & = & \beta^{(p)}_{\mu} + \sum_{k=1}^{M}(1-\omega_{k\mu}), \nonumber \\
T_{\mu}^{z 1}
 &=& \alpha_{z, \mu}^{(\lambda)} + 
 \sum_{k=1}^{M} \left( \sum_{i \in \bm \mu} \delta_{z, \omega_{k\mu}} \delta_{1, s_{ik}} 
 \right),\nonumber \\
 T_{\mu}^{z 0}
& = & \beta_{z, \mu}^{(\lambda)} + 
 \sum_{k=1}^{M} \left( \sum_{i \in \bm \mu} \delta_{z, \omega_{k\mu}} \delta_{0, s_{ik}} 
 \right), 
\label{defvarinP}
\end{eqnarray}
where $\delta$ is Kronecker delta, boldface $\bm \mu$ is the set of neurons in the $\mu$th ensemble, and $z \in \{0,1\}$. 
The variable $G_\mu$ means the number of neurons in the ensemble. The sums in 
$H_{\mu} / \bar{H}_{\mu}$ are the numbers of active/inactive states, respectively. 
The variables $T_{\mu}^{z1}$ and $T_{\mu}^{z0}$ represent coherence between ensemble activity and neuronal activity 
for the same superscript variables (i.e. $T_{\mu}^{11}$ and $T_{\mu}^{00}$), and incoherence (=noise) for the
different superscript variables (i.e. $T_{\mu}^{10}$ and $T_{\mu}^{01}$).
The relation among variables/hyperparameters after integration out of parameters $\{ \bm n, \bm p, \bm \lambda \}$ is depicted in figure \ref{Fig.graphical}(B).

The posterior $P(\bm t, \bm \omega| \bm s)$ can be constructed from joint probability in equation (\ref{P(t, omega, s)}). 
Using this posterior, the membership label $\bm t$ and the ensemble activity $\bm \omega$ 
can be inferred from the variable $\bm s$ or input data, which corresponds to experimental time series data of neuronal activity.

\subsection{The idea of algorithm improvement}
In principle, one can obtain neuronal ensembles and their activities by Bayesian inference with the framework in the last subsection. However, for large scale data the number of possible neuronal states is huge, and direct Bayesian inference is infeasible. 

To cope with computational cost problem, we employ MCMC to evaluate maximum of posterior approximately. 
The idea is originally introduced in the previous study \cite{diana2019bayesian}, and briefly explained in the following.
For Bayesian inference of $\bm t$ and $\bm \omega$, we need conditional probabilities for these variables. 
From equation (\ref{P(t, omega, s)}), the conditional probability for membership label $t_i$ 
for the $i$th neuron is written as
\begin{equation}
P(t_i = \mu | {\bm t}_{\backslash i}, \bm \omega, \bm s) \propto (G_{\mu} + \alpha_{\mu}^{(n)} )
\prod_{z \in \{0,1\}} \frac{B (T_{\mu}^{z1}, T_{\mu}^{z0})}{B (T_{\mu \backslash i}^{z1}, T_{\mu \backslash i}^{z0})},
\label{condt}
\end{equation}
where ${\bm t}_{\backslash i} = \{ t_1, \ldots, t_{i-1}, t_{i+1}, \ldots, t_N \}$. The symbols 
$T_{\mu \backslash i}^{z1}, T_{\mu \backslash i}^{z0}$ are defined by the
replacement of the summation range in equation (\ref{defvarinP}) as $\bm \mu \rightarrow \bm \mu \backslash i$,
where $\bm \mu \backslash i$ the set $\bm \mu$ without $i$.
Note that the backslash notation for exclusion of specific element is used throughout this article.
Similarly, the conditional probability for ensemble activity $\omega_{k \mu}$ for the $\mu$th ensemble at time $k$ is given 
from equation (\ref{P(t, omega, s)}) as
\begin{eqnarray}
P(\omega_{k \mu}=1 | {\bm t}, {\bm \omega}_{\backslash k \mu}, \bm s) &=& \frac{1}{1+ \rho_{k \mu}}, \nonumber \\
 \hspace{2cm} {\rm where}\ \ \rho_{k \mu} &=& \frac{\bar{H}_{\mu} + \beta_{\mu}^{(p)}}{H_{\mu} + \alpha_{\mu}^{(p)}}
 \prod_{z \in \{0,1\}}
 \frac{B(T_{\mu}^{z1}, T_{\mu}^{z0})|_{\omega_{k\mu}=0}}{B(T_{\mu}^{z1}, T_{\mu}^{z0})|_{\omega_{k\mu}=1}}.
\label{condomega}
\end{eqnarray}
Using these probabilities, the variables $\bm t$ and $\bm \omega$ are sampled alternately and iteratively till
they reach convergence. The variables after convergence is regarded as the result of Bayesian inference.

However, in our problem, the number of ensembles $A$ is unknown and to be evaluated as well. Therefore, simple MCMC
algorithm using equations (\ref{condt}) and (\ref{condomega}) should be modified.
For this purpose,  Dirichlet process (DP) \cite{neal2000markov} is introduced and
combined with MCMC in the previous work \cite{diana2019bayesian}.
In DP, when the number of ensembles at current DP step is $A$,  
the ensemble label for each neuron at the next DP step can take the value between $1$ to $A+1$ probabilistically.
Hence, the number of ensembles can increase at the next DP step. By combining DP and MCMC, 
neuronal ensembles both in synthetic activity data and in biological activity data can be inferred successfully
without giving the number of ensembles $A$ \cite{diana2019bayesian}. 

Despite the success of their method, we have one remark. In their method, one needs to start with {\it large} $A$ as initial condition of DP for successful inference. 
By starting small initial $A$, one obtains inappropriate inference result with a few ensembles. 
Since the computational cost is proportional to $A$ for one MCMC step, 
their method still requires large computational cost at early stage of MCMC. 
(See also Algorithm \ref{alg.update}.)

From such background, we propose an improved method to reduce computational cost and to avoid inappropriate result. 
The differences from the previous study are summarized as follows \cite{kimura2020improved}.
\begin{enumerate}
\item{
In our method, when new ensemble is generated in DP, 
{\it multiple} neurons can move to new ensemble {\it simultaneously}
by synchronous update of ensemble label $\bm t$ for all neurons.
In contrast, in the previous study \cite{diana2019bayesian}, 
the update rule of ensemble label is sequential with respect to the neuron label. Accordingly, only one neuron can move to
new ensemble in each update.

}
\item{
We apply the idea of {\it simulated annealing} to transient probability of neurons to new ensemble in DP
for controlling the number of ensembles appropriately.}
\end{enumerate}

Once new ensemble is generated in DP, the fate of new ensemble will be 
different between the previous study and our method.
As mentioned, the ensemble label update is 
sequential in the previous study \cite{diana2019bayesian}, and new ensemble by DP always has only one neuron.
In DP, transient probability is proportional to the number of neurons
in the destination ensemble. (See equation (\ref{DP}).)
Therefore, it is difficult for new ensemble to grow up,
because transient probability to new ensemble is very small for other neurons.
Such small new ensemble will easily be absorbed into other large ensembles during DP iteration.
On the other hand, new ensemble is 
hardly absorbed in our method, because new ensemble can have multiple neurons due to synchronous update of $\bm t$.
Accordingly, transient probability to new ensemble can be relatively large, and
many neurons in new ensemble can remain during DP iteration.
For this reason, new ensemble is hard to vanish and the number of ensembles can easily increase in our method.
In addition, to control the number of ensembles not to increase excessively, 
the idea of simulated annealing should be introduced.
With these ideas, one can infer appropriate ensemble structure by our method
{\it without starting large $A$}, as experimentally shown later. \\

\subsection{Construction of improved algorithm}
The detail of our algorithm is given in the following. 
In the previous work \cite{diana2019bayesian}, 
DP is combined with MCMC algorithm by introducing specific Metropolis-Hastings acceptance 
rule to increase or decrease the number of ensembles, which
can generalize Bayesian inference of $\bm t, \bm \omega$ to arbitrary number of ensembles.
In \cite{diana2019bayesian}, they use the same update rule in equation (\ref{condomega}) for $\bm \omega$,
whereas they use the different rule from (\ref{condt}) for $\bm t$. 

We basically follow their idea. 
In our algorithm, one first updates ensemble activity $\bm \omega$ by equation (\ref{condomega}), then updates ensemble membership label $\bm t$ by the combination rule of MCMC and DP. 
Suppose that there are $A$ ensembles in the intermediate stage of MCMC. In DP, destination ensemble of the $i$th neuron, denoted by $t_{i}^*$, is determined probabilistically by the following distribution,
\begin{equation}
q_{i}(t_{i}^*) = \left\{ 
                \begin{array}{ll} \displaystyle
                \frac{G_{t_{i}^0}^{(\backslash i)}}
                     {q_{\alpha}^{[\gamma]} + N - 1}
                & {\rm for}\ \ t_{i}^* = 1, 2, \ldots, A, \vspace{2mm} \\
                \displaystyle
                \frac{q_{\alpha}^{[\gamma]}} 
                     {q_{\alpha}^{[\gamma]} + N - 1} 
                & {\rm for}\ \ t_{i}^* = A + 1.
                \end{array} 
                \right.
\label{DP}     
\end{equation}
The symbol $G_{t_{i}^0}^{(\backslash i)}$ denotes the number of neurons in the $t_{i}^0$th ensemble, where the $i$th neuron is not counted. Note that $\sum_{\mu=1}^A G_{\mu}^{(\backslash i)} = N-1$ and $q_{i}(t_{i}^*)$ satisfies the property of probability, $\sum_{\mu=1}^{A+1} q_{i}(t_{i}^*) = 1$. 
As seen in equation (\ref{DP}), the parameter $q_{\alpha}^{[\gamma]}$ is proportional to transient probability to the new $(A+1)$th ensemble. In the original work \cite{diana2019bayesian}, the parameter $q_{\alpha}^{[\gamma]}$ is taken to be constant.

Now we apply the idea of simulated annealing to DP. In our method, the parameter at the $\gamma$th MCMC step $q_{\alpha}^{[\gamma]}$ decays exponentially as 
\begin{equation}
q_{\alpha}^{[\gamma]} =  q_{\alpha}^{[0]}
                     e^{- \frac{\gamma}{\tau}},
\label{qdecay}
\end{equation}
where $\tau$ is decay constant. The idea of equation (\ref{qdecay}) is summarized as follows. At early stage of MCMC or small $\gamma$, the number of ensembles $A$ always varies for exploring appropriate $A$, while the change of $A$ is suppressed at late stage of MCMC for convergence.
Note that other function instead of exponential is also applicable. For comparison, we
also apply slowly decaying power-low function in the experiment in the next section.

When new ensemble is generated by the transition of neurons in DP,  
one needs ensemble activity and hyperparameters of new ensemble 
for evaluation of probability ratio in MCMC.
However, there is no prior information of hyperparameters for new ensemble, therefore one can set them arbitrarily.
In our experiment, we set the same hyperparameter values as initially given in MCMC for already-existing ensembles. 
After transition, the activity of new ensemble is set to be the fraction of active neurons,
\begin{eqnarray}
P(\omega_{k, A+1} = 1) = \frac{\sum_{i=1}^N s_{ik}\ \delta_{A+1, t_i^*}}{G_{A+1}}.
\label{newhyperpara}
\end{eqnarray}
The activity of new ensemble before transition is set to be random, 
because it does not exist before transition.
In addition, it should be noted that the number of ensembles can decrease, because an already-existing ensemble is deleted
when it becomes empty (=no neuron) after MCMC update. 

For computing acceptance rate of new membership label in MCMC,
let us consider the case that the membership labels $\bm t^0 = \{t_1^0, t_2^0, \ldots, t_N^0 \}$ may be updated to new ones $\bm t^* = \{ t_1^*, t_2^*, \ldots, t_N^* \}$. The probability ratio between $\bm t^0$ and $\bm t^*$ is calculated from equation (\ref{P(t, omega, s)}), 
\begin{eqnarray}
\frac{P(\bm t^*, \bm \omega, \bm s)}
           {P(\bm t^0, \bm \omega, \bm s)} 
& = & \frac{ 
            \prod_{\mu = 1}^{A+1}
            \Gamma(\alpha_{\mu}^{(n)} + G_{\mu})
            \left. \left\{
              \prod_{z = \{0, 1\}}
              \it{B}({T}_{\mu}^{z1},{T}_{\mu}^{z0})
            \right\}
            \right|_{\bm t = \bm t^*}
           }
           { 
            \prod_{\mu = 1}^{A+1}
            \Gamma(\alpha_{\mu}^{(n)} + G_{\mu})
            \left. \left\{
              \prod_{z = \{0, 1\}}
              \it{B}({T}_{\mu}^{z1},{T}_{\mu}^{z0})
            \right\}
            \right|_{{\bm t} = {\bm t}^0}
           }
           \nonumber \\ 
& &   \cdot
      \frac{ \left. \it{B}(H_{A+1}, \bar{H}_{{A+1}}) \right|_{\bm t = \bm t^*} }
           { \left. \it{B}(H_{A+1}, \bar{H}_{{A+1}}) \right|_{\bm t = \bm t^0} }. 
\label{transient rate}
\end{eqnarray}
If there is no transient neuron to new ensemble, the factors for the new $(A+1)$th ensemble in the denominator and the numerator cancel out. 

Next, for Metropolis-Hastings update rule, one also needs to define proposal distribution from the $t_{i}^{0}$th ensemble to the $t_{i}^{*}$th, $Q_{i}(t_{i}^{\ast} | t_{i}^{0})$, and its reverse process $Q_i(t_{i}^{0} | t_{i}^{\ast})$ for the $i$th neuron. From detailed balance condition, they are calculated as
\begin{eqnarray}
Q_{i}(t_{i}^{\ast} | t_{i}^{0}) & = & \left\{
                                     \begin{array}{ll}
                                     \displaystyle \frac{G_{t_{i}^0}^{(\backslash i)}}{q_{\alpha}^{[\gamma]}+ N - 1}
                                     & {\rm for} \ \ t_{i}^* = 1, 2, \ldots, A, \vspace{1mm} \\
                                     \displaystyle \frac{q_{\alpha}^{[\gamma]}}{q_{\alpha}^{[\gamma]}+ N - 1}
                                     & {\rm for} \ \ t_{i}^* = A + 1,    
                                     \end{array} \label{proposal}
                                     \right. \\
Q_{i}(t_{i}^{0} | t_{i}^*) & = & \frac{G_{t_{i}^{0}}^{(\backslash i)}}{N-1},
\end{eqnarray}
for the $\gamma$th MCMC step. Note that r.h.s. of equation (\ref{proposal}) is the same as that of 
equation (\ref{DP}). When multiple neurons move simultaneously, the product of the proposal distribution (\ref{proposal}) for all transient neurons must be considered. 

Now we arrive at the stage to construct the rule for updating membership label $\bm t$ in MCMC.
In our algorithm, we first determine the new membership label $\bm t^*$ for all neurons 
by DP in equation (\ref{DP}). Next, we compute
the acceptance rate $a(t^*_i, t^0_i)$ from the membership label $t^0_i$ to $t^*_i$ 
for the $i$th neuron in standard Metropolis-Hastings rule as
\begin{eqnarray}
\label{acceptance rate}  
a(t^*_i, t^0_i) &=& \left\{
                     \begin{array}{l}
                      \displaystyle {\rm min}
				      \left\lbrace	
                      1,\ 
                      \frac{P(t^*_i, \bm t^{0}_{\backslash i}, \bm \omega, \bm s)}
                           {P(\bm t^{0}, \bm \omega, \bm s)}  
					  \frac{Q(t^{0}_i | t^{*}_i)}
						   {Q(t^{*}_i | t^{0}_i)}
                      \right\rbrace \\
                      \hspace{6cm} {\rm for} \ \ t_{i}^* = 1, 2, \ldots, A, \\ \\
                      \displaystyle {\rm min}
				      \left\lbrace	
                      1,\ 
                      \frac{P(\bm t^*_{A+1}, \bm t^{0}_{\backslash A+1}, \bm \omega, \bm s)}
                           {P(\bm t^{0}, \bm \omega, \bm s)}  
					  \frac{Q(\bm t^{0} | \bm t^*_{A+1}, \bm t^{0}_{\backslash A+1})}
						   {Q(\bm t^{*}_{A+1}, \bm t^{0}_{\backslash A+1} | \bm t^{0})}
                      \right\rbrace \\
                      \hspace{7cm} {\rm for} \ \ t_{i}^* = A+1, \vspace{1mm} \\
                     \end{array}  
                    \right.
\end{eqnarray}
where equation (\ref{transient rate}) is used for computation of probability ratio. 
The sets of ensemble labels regarding the new ($A+1$)th ensemble are defined by
\begin{eqnarray}                                       
 \bm t_{A+1}^* &=& \{\ t_j^*\ |\ j\in\{1.\ldots,N\}, t_j^* = A+1 \}, \nonumber \\
 \bm t_{\backslash A+1}^* &=& \{\ t_j^*\ |\ j\in\{1.\ldots,N\}, t_j^* \ne A+1 \}, \nonumber \\
 \bm t_{A+1}^0 &=& \{\ t_j^0\ |\ j\in\{1.\ldots,N\}, t_j^0 = A+1 \}, \nonumber \\
 \bm t_{\backslash A+1}^0 &=& \{\ t_j^0\ |\ j\in\{1.\ldots,N\}, t_j^0 \ne A+1 \}.
\end{eqnarray}
Note that the proposal probability can be decomposed for computation of its probability ratio in equation
(\ref{acceptance rate}).
\begin{eqnarray} 
 Q (\bm t^0 | \bm t^*) = \prod_{i=1}^N Q_i (t_i^0 | t_i^*),\ \ Q (\bm t^* | \bm t^0) = \prod_{i=1}^N Q_i (t_i^* | t_i^0).
\end{eqnarray}
Using this acceptance rule in (\ref{acceptance rate}), 
the new membership label $t^*_i$ is accepted or rejected probabilistically in MCMC.
In our method, one first determines destination ensemble $\bm t^*$ synchronously, where
multiple neurons can have membership label of new ensemble as destination. Then one computes acceptance rate for each neuron. 
In this way, our method enables multiple neurons to move to new ensemble simultaneously.

Finally, hyperparameters of priors must also be updated. For hyperparameter update, learning rate $\varepsilon^{[\gamma]}$ is introduced to control the influence by simulated annealing, where $\gamma$ is MCMC step as in equation (\ref{qdecay}). 
In our method, we choose sigmoid function for learning rate because it is bounded and smooth,
\begin{equation}
\varepsilon^{[\gamma]} = \frac{1}{1 + e^{- \frac{\gamma}{\tau}}},
\label{epsilonupdate}
\end{equation}
where decay constant $\tau$ is the same as in equation (\ref{qdecay}). 
By following the original work \cite{diana2019bayesian}, hyperparameters are updated with learning rate $\varepsilon^{[\gamma]}$ as 
\begin{eqnarray}
\tilde{\alpha}_{\mu}^{(p)} & = & 
\alpha_{\mu}^{(p)} + \varepsilon^{[\gamma]}
                     \left(
                     \sum_{k=1}^{M} \omega_{k \mu}
                     \right),\nonumber \\
\tilde{\beta}_{\mu}^{(p)} & = & 
\beta_{\mu}^{(p)} + \varepsilon^{[\gamma]}
                    \left(
                    \sum_{k=1}^{M} (1 - \omega_{k \mu})
                    \right), \nonumber \\
\tilde{\alpha}_{z, \mu}^{(\lambda)} & = & 
\alpha_{z, \mu}^{(\lambda)} + \varepsilon^{[\gamma]} 
                              \left(
                              \sum_{k=1}^{M}
                              \left(
                              \sum_{i \in \bm \mu}
                              \delta_{z, \omega_{k \mu}}
                              \delta_{1, s_{i k}}
                              \right)
                              \right), \nonumber \\
\tilde{\beta}_{z, \mu}^{(\lambda)} & = & 
\beta_{z, \mu}^{(\lambda)} + \varepsilon^{[\gamma]} 
                             \left(
                             \sum_{k=1}^{M}
                             \left(
                             \sum_{i \in \bm \mu}
                             \delta_{z, \omega_{k \mu}}
                             \delta_{0, s_{i k}}
                             \right)
                             \right), \nonumber \\
\tilde{\alpha}_{\mu}^{(n)} & = & 
\alpha_{\mu}^{(n)} + \varepsilon^{[\gamma]} G_{\mu},
\label{hyperpara-update}
\end{eqnarray}
where tilde means updated hyperparameter. 

Note that the learning rate (\ref{epsilonupdate}) is closely related 
with the parameter $q_{\alpha}^{[\gamma]}$ in equation (\ref{qdecay}) in our method.
At early stage of MCMC, small learning rate will suppress the change of hyperparameters, while the number of ensembles $A$ varies frequently as mentioned.
Hence, it is natural to choose the function of learning rate $\varepsilon^{[\gamma]}$ according to the function of $q_{\alpha}^{[\gamma]}$. 
As mentioned, for comparison with exponential decay of $q_{\alpha}^{[\gamma]}$,
we also apply power-law decay of $q_{\alpha}^{[\gamma]}$ in the experiment, where 
we use another learning rate function including power-law decay factor.

To summarize, our algorithm is expressed as the pseudo code in Algorithm \ref{alg.update}. 

\begin{algorithm}[H]                 
\caption{Inference of ensembles, ensemble activities, and the number of ensembles}         
\label{alg.update}                          
\begin{algorithmic}                  
\STATE{initialize $\bm \omega$ and $\bm t$} 
\WHILE{the number of ensembles $A$ converges}
\FOR{each ensemble $\mu \in \{1, 2, \ldots, A\}$, $k \in \{1, 2, \ldots, M\} $}
\STATE{draw $\omega_{k \mu} \sim 
             P(\omega_{k \mu}=1|\bm t, \bm \omega_{\backslash k \mu}, \bm s)$} in equation (\ref{condomega})
\ENDFOR
\FOR{each neuron $i \in \{1, 2, \ldots, N\}$}
\STATE{draw destination ensemble $t_i^{\ast} \sim q(t_i^{\ast})$ in equation (\ref{DP})}
\IF{$t_i^{\ast} = A+1$}
\STATE{$A \rightarrow A+1$}
\STATE{give new hyperparameters}
\ENDIF
\ENDFOR
\FOR{each neuron $i \in \{1, 2, \ldots, N\}$}
\STATE{accept or reject $t_i^*$ according to acceptance rate $a(t_i^{*}, t_i^{0})$ 
       in equation (\ref{acceptance rate})}
\ENDFOR
\FOR{each ensemble $\mu \in \{1, 2, \ldots, A\}$}
\IF{$G_{\mu} = 0$}
\STATE{delete the $\mu$th ensemble and its hyperparameters}
\ENDIF
\ENDFOR
\STATE{update hyperparameters as equation (\ref{hyperpara-update})}
\ENDWHILE
\end{algorithmic}
\end{algorithm}

\section{Experiment 1: application to synthetic data}
\subsection{Profile of synthetic data}
For validation of our method, we conduct numerical experiment of neuronal ensemble inference using synthetic data.
 In our experiment, synthetic data with ground-truth ensembles is generated by Algorithm \ref{alg.generate}. Then, ensembles are inferred from this synthetic data by Algorithm \ref{alg.update}, whose result is compared with the ground-truth ensembles. 

In our generative model, neuronal activities are closely related to ensemble activities by equation (\ref{conductivity}), and the relation between them is characterized by conditional activity rate $\bm \lambda$. To generate synthetic data, 
all neurons are divided into ground-truth ensembles first. Then, ensemble activity $\bm \omega$ is generated for each ensemble by activity parameter $\bm p$. Finally, activity of each neuron $\bm s$ is determined using ensemble activity $\bm \omega$ and conditional activity rate $\bm \lambda$.
The algorithm of synthetic data generation is summarized as the pseudo code in Algorithm \ref{alg.generate}. 
See also the original work for the detail \cite{diana2019bayesian}.

\begin{algorithm}[H]                      
\caption{Generation of synthetic neuronal activity data}                             
\begin{algorithmic}                  
\STATE{set all $\bm \omega$ and $\bm s$ to be 0}
\FOR{each ensemble $\mu \in \{1, 2, \ldots, A\}$, $k \in \{1, 2, \ldots, M\} $}
\STATE{draw $\omega_{k\mu} \sim P(\omega_{k\mu})$}
\ENDFOR
\FOR{each neuron $i \in \{1, 2, \ldots, N\}$}
\STATE{deal the $i$th neuron to an ensemble}
\ENDFOR
\FOR{each neuron $i \in \{1, 2, \ldots, N\}$, $k \in \{1, 2, \ldots, M\}$}
\IF{$\omega_{kt_{i}} = 1$}
\STATE{draw $s_{ik} \sim P(s_{ik}|\omega_{kt_{i}}=1)$ in equation (\ref{conductivity})}
\ELSE
\STATE{draw $s_{ik} \sim P(s_{ik}|\omega_{kt_{i}}=0)$ in equation (\ref{conductivity})}
\ENDIF
\ENDFOR
\end{algorithmic}
\label{alg.generate}
\end{algorithm}

\begin{figure}
\begin{center}
	\includegraphics[scale=0.8]{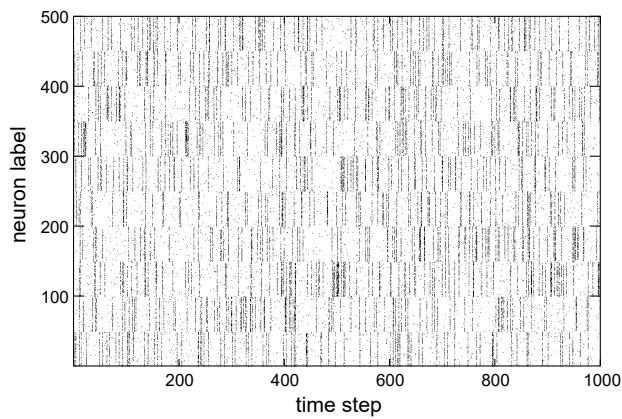}
\caption{An example of synthetic data of neuronal activity by the Algorithm \ref{alg.update}. It has $500$ neurons and
         $10$ ground-truth ensembles with $1000$ time steps. (black=active neuron, white=inactive neuron)}
\label{Fig.gendata}
\end{center}
\end{figure}

In figure \ref{Fig.gendata}, an example of activity $\bm s$ by Algorithm \ref{alg.generate}
is illustrated. In this example, there are 10 ground-truth ensembles and 500 neurons, where each ensemble has 50 neurons equally. The vertical axis represents neuron label, which is sorted by neuronal membership label $\bm t$. 
The 10 ensemble structure can be seen clearly, however such structure cannot be recognized easily if neuron labels are randomly permuted. The values of parameters for synthetic data are given in Table \ref{Table.generation condition}. All ensembles/neurons are generated with the same ensemble activity rate $\bm p$ and conditional activity rate $\bm \lambda$. 

\begin{table}
\caption{The parameters for synthetic data generation}
\begin{center}
\begin{tabular}{c|c} \hline 
parameter & value \\ \hline
the number of neurons & $N=500$ \\
the number of ensembles & $A=10$ \\
ensemble activity rate & $p_{\mu}=0.1\ \ (\forall \mu)$ \\
conditional activity rate \ & \
$\lambda_{\mu}(0)=0.01$,  $\lambda_{\mu}(1)=0.6$ \ $(\forall \mu)$\\ \hline 
\end{tabular}
\end{center}
\label{Table.generation condition}
\end{table}

\begin{figure}
\begin{center}
	\includegraphics[scale=0.8]{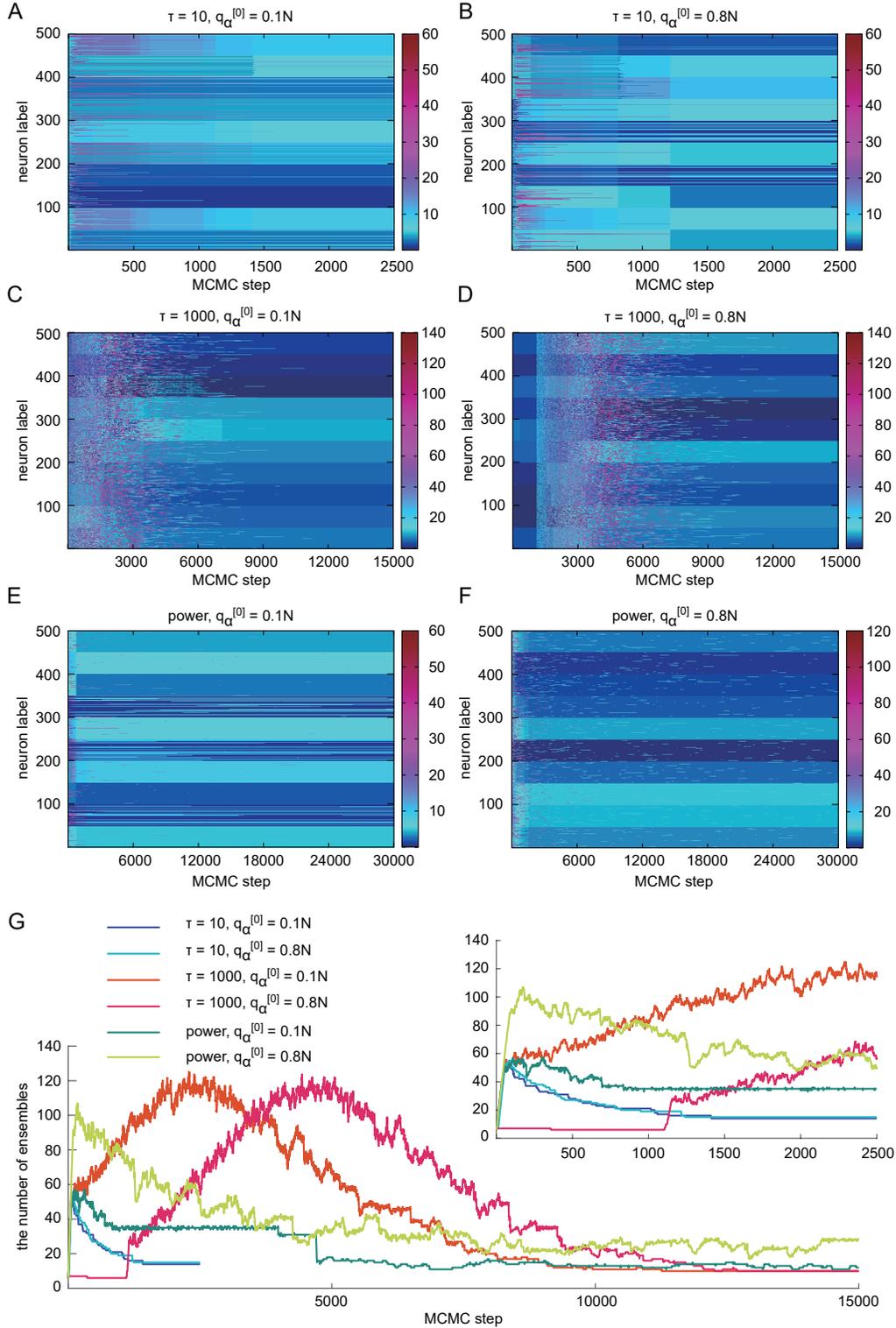}
\caption{The inference result of neuronal ensembles for the data in figure \ref{Fig.gendata} 
  by the Algorithm \ref{alg.update}:
	     (A) Behavior of ensemble membership label for exponential $q_{\alpha}^{[\gamma]}$ with $\tau=10, q_{\alpha}^{[0]}=0.1N$.
	     The color in the heat map represents ensemble number.
	     (B) exponential $q_{\alpha}^{[\gamma]}$ with $\tau=10, q_{\alpha}^{[0]}=0.8N$
	     (C) exponential $q_{\alpha}^{[\gamma]}$ with $\tau=1000, q_{\alpha}^{[0]}=0.1N$
	     (D) exponential $q_{\alpha}^{[\gamma]}$ with $\tau=1000, q_{\alpha}^{[0]}=0.8N$
	     (E) power-law $q_{\alpha}^{[\gamma]}$ with $q_{\alpha}^{[0]}=0.1N$
	     (F) power-law $q_{\alpha}^{[\gamma]}$ with $q_{\alpha}^{[0]}=0.8N$
         (G) Behavior of the number of ensembles during MCMC in six cases.}
\label{Fig.result}
\end{center}
\end{figure}

\subsection{Experiment of ensemble inference}
\label{Sec:NV}

For ensemble inference, the synthetic data in figure \ref{Fig.gendata} is used as input activity $\bm s$. 
In our experiment of inference, ensemble activity $\bm \omega$, ensemble membership label $\bm t$, and hyperparameters are updated  iteratively till the number of transient neurons to other ensemble in DP becomes sufficiently small.
 In this experiment, we set initial number of ensembles $A=5$, initial transient parameter $q_{\alpha}^{[0]}=0.1N$ or $0.8N$, and hyperparameters $\alpha_{\mu}^{(p)}=100, \beta_{\mu}^{(p)}=100, \alpha_{z,\mu}^{(\lambda)}=100, \beta_{z,\mu}^{(\lambda)}=100, \alpha_{\mu}^{(n)}=100$ for all $\mu,z$. 
For the parameter $q_{\alpha}^{[\gamma]}$, we use the exponential decay rule in equation (\ref{qdecay}) 
with decay constant $\tau=10$ or $1000$ and the power-law decay rule $q_{\alpha}^{[\gamma]} =q_{\alpha}^{[0]} / \gamma$.
The function of the learning rate $\varepsilon^{[\gamma]}$ is chosen as in equation (\ref{epsilonupdate})
for the exponential $q_{\alpha}^{[\gamma]}$, 
and $\varepsilon^{[\gamma]} = 1 / (1+\gamma^{-1})$ for the power-law $q_{\alpha}^{[\gamma]}$.
At initialization step, initial membership label is randomly assigned to each neuron uniformly within the range between 1 to $A$. 
 
Dynamical behaviors of membership labels in MCMC with different $\tau$ and $q_{\alpha}^{[0]}$ 
are shown in figure \ref{Fig.result}(A)-(F) by the heat map.  
We conduct 2500 MCMC updates for exponential $q_{\alpha}^{[\gamma]}$ with $\tau=10$, 
$15000$ updates for exponential $q_{\alpha}^{[\gamma]}$ with $\tau=1000$,
and $30000$ updates for power-law $q_{\alpha}^{[\gamma]}$.
Ensemble numbers \{$1, 2, \ldots, A$\} are identified by the colors in the heat map. 
In figure \ref{Fig.result}(G), dynamical behavior of the number of ensembles during MCMC is depicted. 

In all six cases, the ground-truth ensemble structure can be observed at late stage of MCMC.
In the exponential $q_{\alpha}^{[\gamma]}$, 
some neurons are not classified into correct ensemble even at late stage for smaller $\tau$,
while for larger $\tau$ the number of such incorrect neurons is very small.
Namely, larger $\tau$ is favorable for appropriate ensemble inference.
However, for larger $\tau$, the number of ensembles becomes over 100 at intermediate stage of MCMC,
then decreases to the correct value.
This means that most of the ensembles at intermediate stage do not contribute to 
the final result. Nevertheless, much computational cost is required due to too many ensembles at intermediate stage.
Therefore, appropriate value of $\tau$ must be chosen for practical use.
In contrast, the parameter $q_{\alpha}^{[0]}$ does not affect the inference result significantly.
Next, in the power-law $q_{\alpha}^{[\gamma]}$, the convergence of ensemble label is very slow and
many neurons are classified into incorrect ensembles even at late stage of MCMC.
Therefore, slow annealing schedule like 
power-law is not appropriate in our algorithm. We use exponential decay of $q_{\alpha}^{[\gamma]}$ 
in equation (\ref{qdecay}) hereafter.

We give some remarks on our proposed algorithm. First, ground-truth 10 ensemble structure 
can be obtained even under large initial number of ensembles, as required in the original algorithm.
Second, ground-truth ensemble activities ${\bm \omega}$ can also be inferred almost perfectly.
Third, if conditional activity rate $\bm \lambda$, which controls coherence or noise, is varied, the ensemble inference becomes easy/hard. Even under hard condition or noisy case, almost correct ensemble structure can still be obtained and as much noise can be removed as possible. Finally, even when the sizes of ground-truth ensembles are not equal unlike figure \ref{Fig.gendata}, correct ensembles structure can be inferred.

\subsection{Comparison with the original algorithm}
\label{Sec:CompTime}

\begin{figure}
\begin{center}
	\includegraphics[scale=0.9]{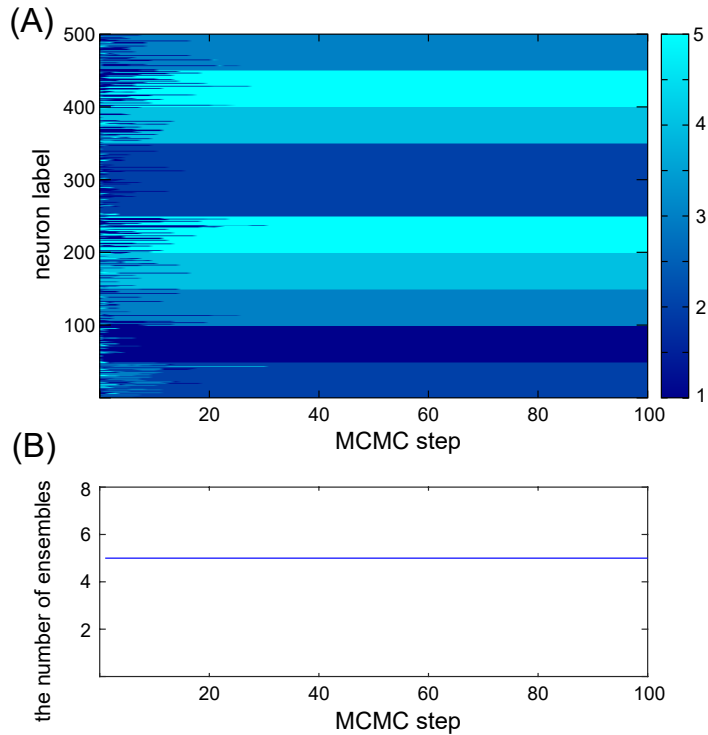}
\caption{Dynamical behavior of the original algorithm in Ref.[12] in MCMC: (A)
Behavior of ensemble membership label. The color in the heat map represents ensemble number.
(B) Behavior of the number of ensembles.}
\label{Fig.original}
\end{center}
\end{figure}

\begin{figure}
\begin{center}
	\includegraphics[scale=1.1]{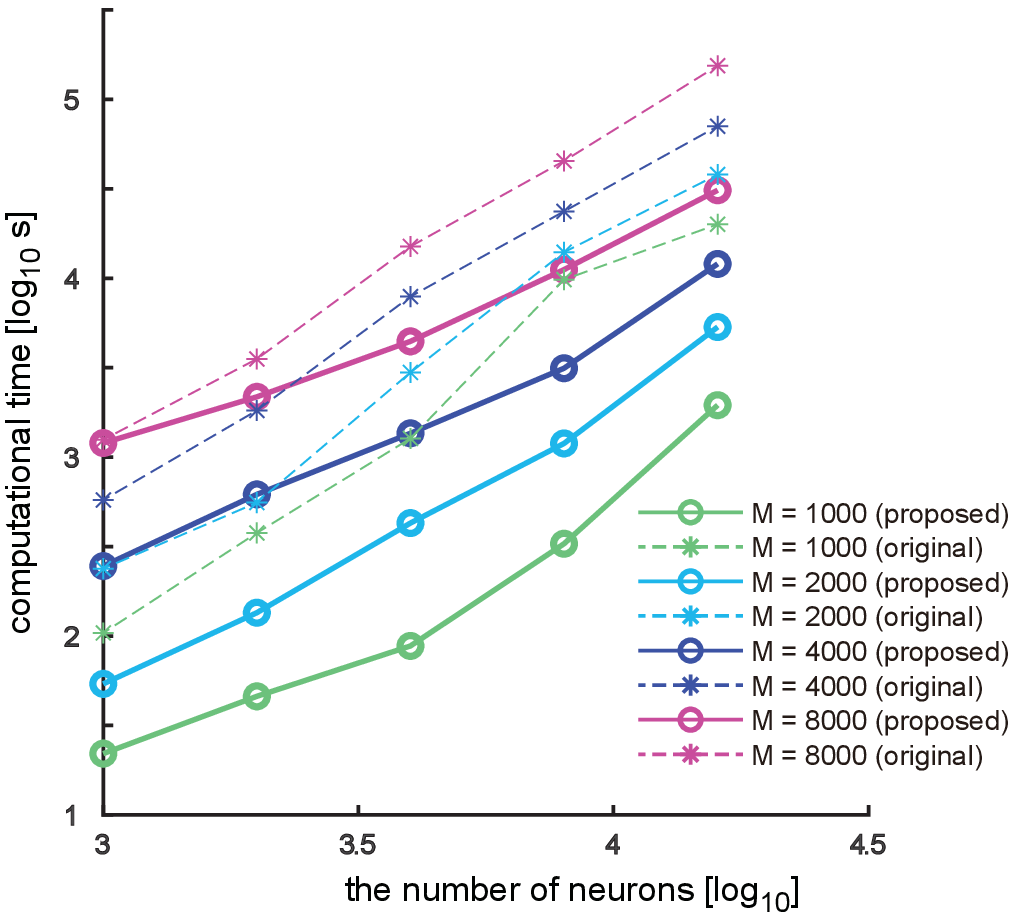}
\caption{The comparison of computational time between our proposed algorithm and the original in reference \cite{diana2019bayesian}.}
\label{Fig.result2}
\end{center}
\end{figure}
We also apply the original algorithm in \cite{diana2019bayesian} to the same synthetic data in figure 
\ref{Fig.gendata}. 
In \cite{diana2019bayesian}, they recommend the number of initial ensembles is chosen to be $N/2$, while
in this experiment we prepare 5 initial ensembles for comparison with ours.
The same parameter values are used as in subsection \ref{Sec:NV} for common parameters in two algorithms.
The result is shown in figure \ref{Fig.original}.
The correct boundaries between ensembles are observed, however
the number of ensembles does not increase from the initial value 5. 
Accordingly, some ground-truth ensembles are merged in the final result.
In the original algorithm, only single neuron can move to new ensemble.
With such transition rule, the number of ensembles is hard to increase,
because new ensemble having single neuron is easily absorbed to other large ensemble during MCMC.

We also compare computational times by our proposed algorithm and the original.
In this experiment, the numbers of neurons and time steps in the synthetic data are varied.
For the synthetic data, the parameter values excepting the number of neurons 
are the same as in Table \ref{Table.generation condition}.
The experiment is executed by the workstation with 2 CPUs
(Xeon Gold 6238R 2.2GHz with 56 cores in each CPU), and the size of RAM is 256GB. Program code of our proposed algorithm
is written by C++. The C++ code of the original algorithm is available at author's GitHub.
In the original algorithm the initial number of ensembles is $N/2$ as recommended,
while in our proposed algorithm the initial number of ensembles is $5$, $\tau=10$, and $q=0.1N$.
The same parameter values as in subsection \ref{Sec:NV} are used for other parameters.

The computational times for 1000 MCMC steps by two algorithms are summarized in figure \ref{Fig.result2}.
In all experimental settings, the inference result gives almost correct blockwise ensemble structure.
The time of our proposed algorithm is much smaller than the original, which means
our improvement for reduction of computational cost works efficiently.
The original algorithm needs much cost at early stage of MCMC
for activity inference of too many ensembles.
However, the number of ensembles decreases significantly at late stage, 
therefore inference of ensemble activity at early stage is wasteful. In contrast,
MCMC in ours can be conducted with small initial number of ensembles. 
Consequently, computational cost of our proposed algorithm becomes much smaller.

\section{Experiment 2: Application to real activity data}
\label{Sec:ApplyReal}

\subsection{Profile of real activity data}
We move on to the application to a real neuronal dataset, 
an open-access neural activity dataset from the Collaborative Research in Computational Neuroscience (CRCNS) repository 
\cite{TS2009}. 
The dataset we make use of is the alm-1 dataset, which contains single unit spike timings recorded 
by silicon probes from the anterior lateral motor cortex (ALM) \cite{li2015motor}. 
The total number of neurons is 1408 neurons in 19 mice. 
In the experiment for data acquisition, the mouse performs the discrimination task for object location. 
The mouse discriminates the location of a pole using its whiskers and reports its choice with licking. 
The whole experiment consists of three epochs in the following order.
\begin{enumerate}
\item{
The pole is presented at one of two possible positions, anterior or posterior pole position. 
The mouse contacts the pole with their whisker to recognize its position.
}
\item{
The pole is retracted away from the mouse. The mouse has to wait without licking till a "go cue" is presented. 
}
\item{
After go cue, the mouse reports its choice with licking one of two lickports. If the pole position is posterior, 
the mouse is rewarded by licking the left port, and if the pole position is anterior, the mouse is rewarded by 
licking the right port.
}
\end{enumerate}
In the original paper, these epochs are called Sample, Delay, and Response epochs, respectively. The go cue sounds 
at the start of Response epoch so that the mouse can recognize the change of epochs. As a reward for learning, a small drop of water is given if the mouse licks the correct lickport. The mouse collects sensory information during Sample epoch and maintains a memory of pole position or motor choice during Delay epoch. The ALM is involved in planning licking direction, and neuronal activity of the ALM is recorded to observe its relation with mouse's action.

\subsection{Data processing}

\begin{figure}
\begin{center}
	\begin{center}
	\includegraphics[scale=1]{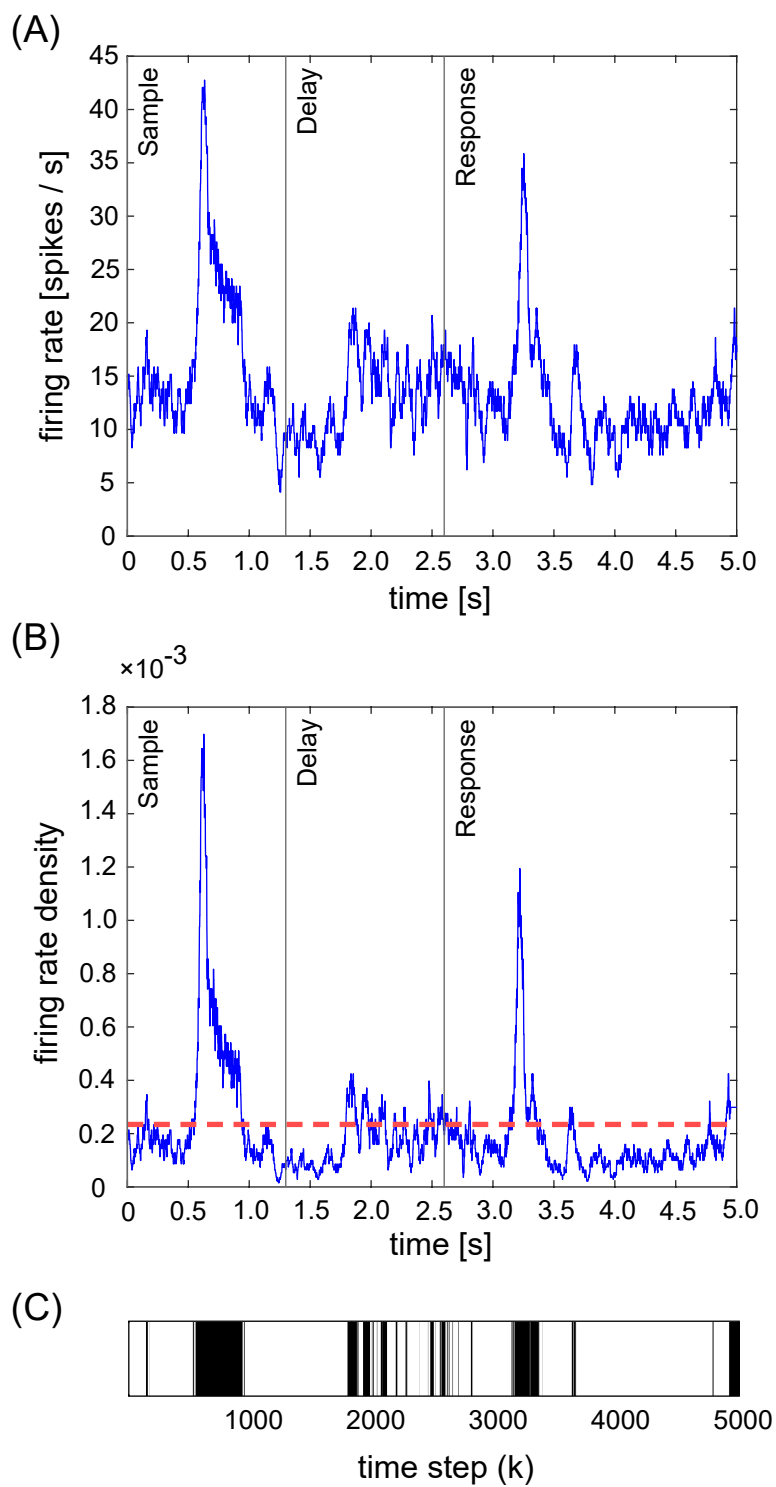}
	\end{center}
	\caption{An example of data processing: 
	         (A) The number of firing events per second for a neuron. 
	         (B) Firing rate (solid) 
	             and kurtosis (horizontal broken). 
	         (C) Binary activity data.} 
\label{Fig.convert}
\end{center}
\end{figure}

The original spike timing data includes multiple trials of pole location detection.
In one trial, spike timing data with a temporal resolution of 0.1 millisecond is stored for 5 seconds. 
This data set is very sparse, because 0.1 millisecond is too short to describe neuronal activity. 
It cannot be used for ensemble inference directly.
Hence, to capture dynamical behavior of neuronal activity appropriately, 
we convert spike timing data to firing rate, 
then obtain non-sparse binary data by thresholding.

First, the data is coarse-grained to have 10 millisecond single time step,
and the number of firing events is counted in each coarse-grained time step.
We also use smoothing to avoid too sparse data:
the activity at every 10 millisecond time step is determined by the sum of the number of firing events between
the current time step and after 50 millisecond.
The activity is averaged over all trials. 
In figure \ref{Fig.convert}(A), the average number of firing events per second is illustrated
for a single neuron.

 Next, the activity data in figure \ref{Fig.convert}(A) is normalized 
 so that the sum of activity data over all time steps
 becomes unity. In some sense, the activity data is converted to probability distribution.
 The normalized distribution in figure \ref{Fig.convert}(B) is called firing rate density. 
 Then, the kurtosis of this distribution is calculated, and the firing rate density is binarized 
 using this kurtosis as threshold.
 In figure \ref{Fig.convert}(B), solid line represents firing rate density, and horizontal broken line represents
 kurtosis of this distribution.
 After binarization, the data in figure \ref{Fig.convert}(C) is obtained, 
 where black/white regions describe active/inactive states, respectively.
 From this figure, our data conversion gives appropriate binarized neuronal activity from spike timing data. 
 
\subsection{Result}
\begin{figure}
	\begin{center}
	\includegraphics[scale=0.9]{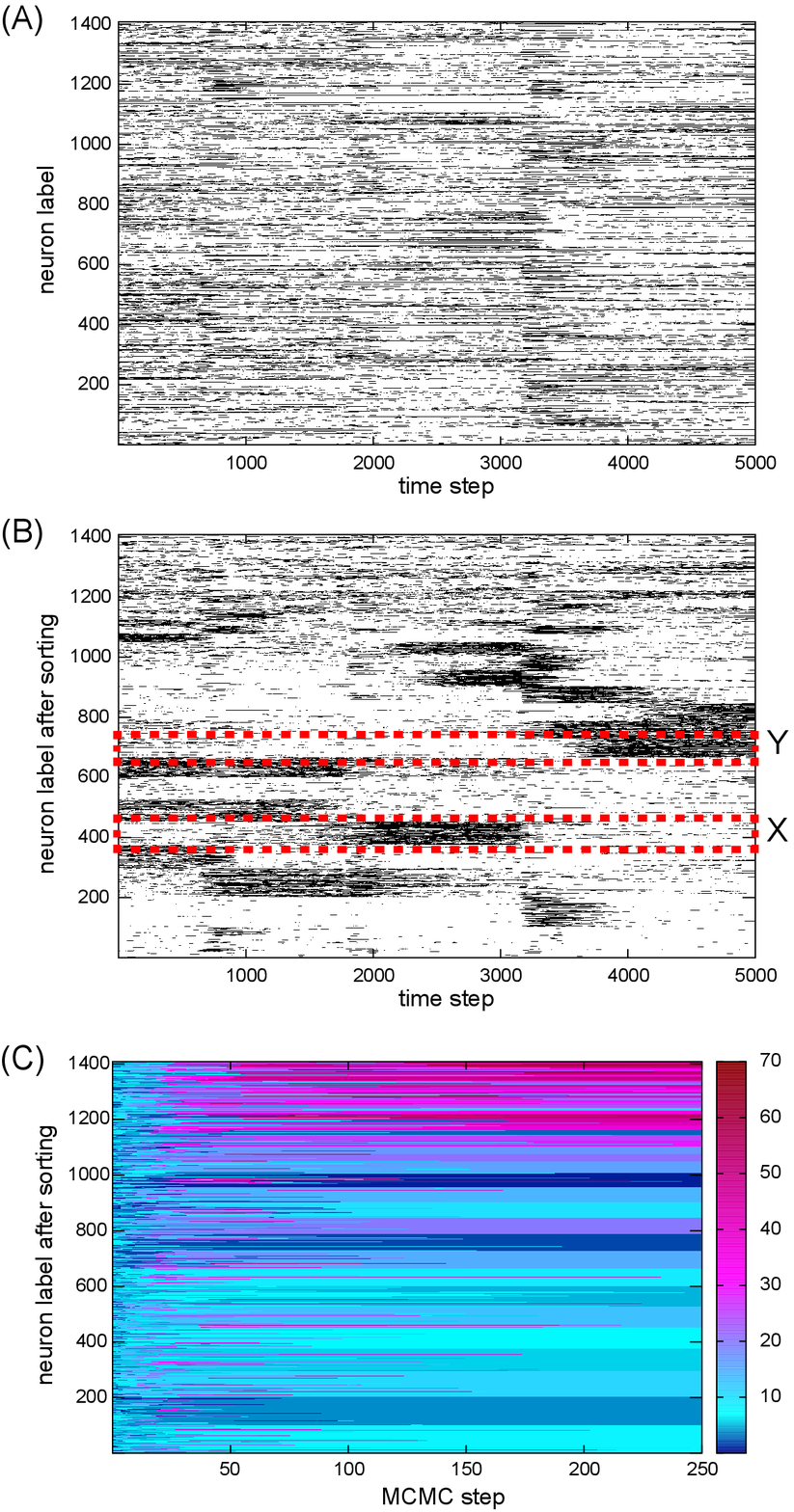}
	\caption{Typical inference result and dynamical behavior 
	         in MCMC: 
	         (A) The binary activity data of all 1408 neurons.
	             Black/white points describe active/inactive neurons, respectively.
	         (B) Activity data of all neurons after sorting 
	             neuron labels.
	         (C) Behavior of ensemble membership label during MCMC. 
	             The color in the heat map represents ensemble number.} 
\label{Fig.real_result}
	\end{center}
\end{figure}

We apply our algorithm to binary data after data preprocessing.
The binary neuronal activity ($N = 1408$, $M = 5000$) by our method is illustrated in figure \ref{Fig.real_result}(A), where ensemble structure cannot be easily recognized.
For inference of ensembles and their activities, our algorithm is executed 
using the data in figure \ref{Fig.real_result}(A) as input activity $\bm s$. 
In our numerical analysis, we conduct 250 iterations in MCMC, where 
the parameters are set as follows: initial number of ensembles $A = 5$, decay constant $\tau = 10$, initial transient parameter $q_{\alpha}^{[0]} = 0.1N$, and hyperparameters $\alpha_{\mu}^{(p)} = 100$, $\beta_{\mu}^{(p)} = 100$, $\alpha_{z, \mu}^{(\lambda)} = 100$, $\beta_{z, \mu}^{(\lambda)} = 100$, $\alpha_{\mu}^{(n)} = 100$ for all $\mu$, $z$. At initialization step, initial ensemble membership label to each neuron is assigned randomly and uniformly within the range between $1$ to $A$. 

\begin{figure}
	\begin{center}
	\includegraphics[scale=1]{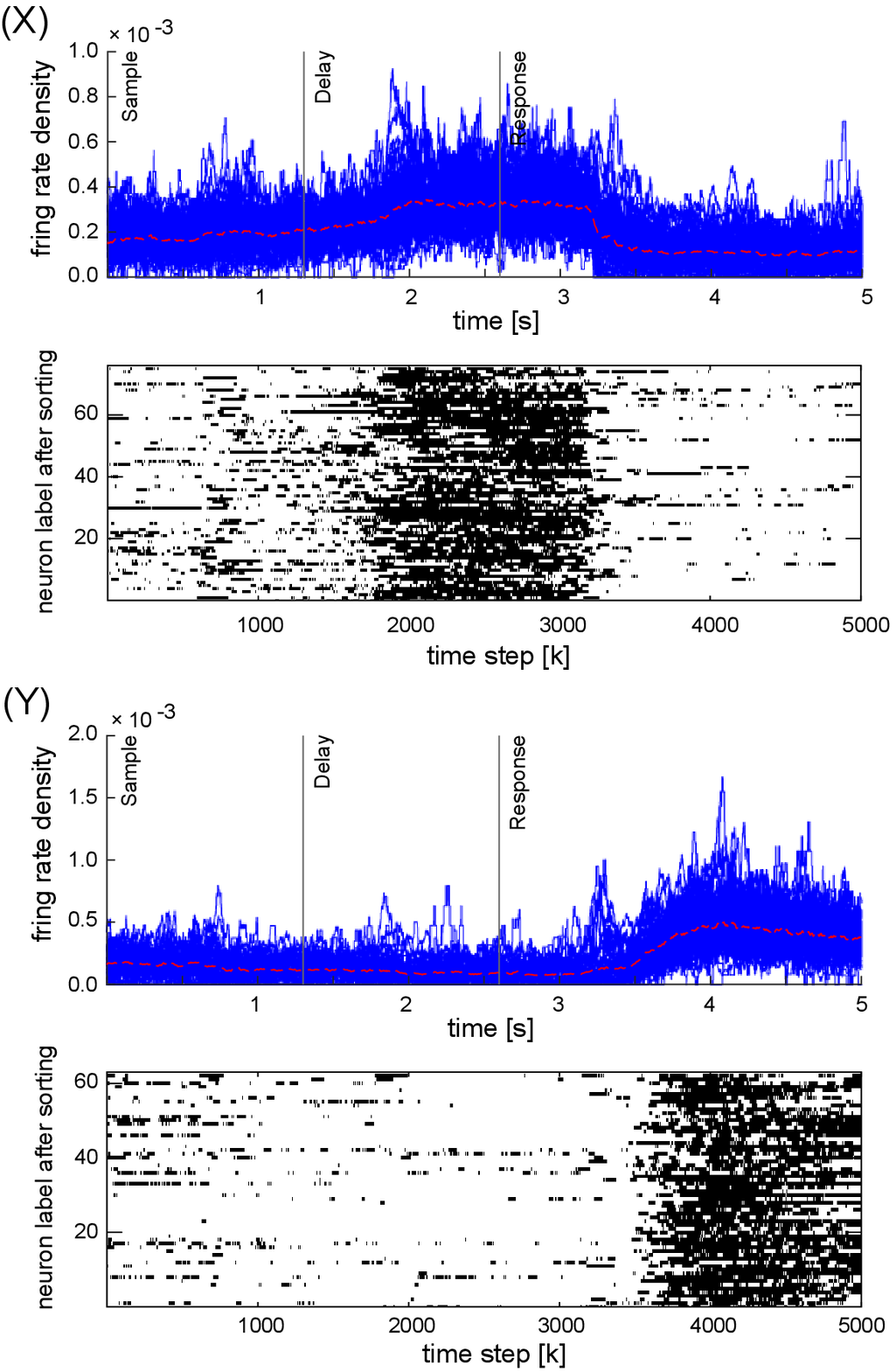}
	\caption{Activities of neurons in the ensembles X/Y in figure \ref{Fig.real_result}: 
	         In the upper figures in ensembles X and Y,
	         thin solid line indicates firing rate density, and 
	         red dotted line is average density.
	         In the lower figures, binary activities are shown.
	         }
\label{Fig.real_part}
    \end{center}
\end{figure}


A typical result of inference is shown in figure \ref{Fig.real_result}(B), where 
neuron labels are sorted using the result of ensemble inference. 
The computational time by the same workstation as in subsection \ref{Sec:CompTime} is 187[s].
One can observe synchronous activity of neurons in the same ensemble.
In figure \ref{Fig.real_result}(C), dynamical behavior of membership labels in MCMC is expressed by the heat map. 
The final number of ensembles is 62 and the sizes of many ensembles are small. 
Most of neurons belong to one of about 20 ensembles, whose sizes are relatively large.
From these figures, ensemble membership labels converge after $200$ MCMC steps.
It is also verified that similar number of ensembles and clear ensemble structure 
are obtained again, even if our algorithm is applied with different initial membership labels.

In figure \ref{Fig.real_part}, activities of neurons in several ensembles are expressed.
The activities in figure \ref{Fig.real_part}(X)/\ref{Fig.real_part}(Y) 
corresponds to the ensembles X/Y in figure \ref{Fig.real_result}, respectively.
In the upper figures in both ensembles, thin solid line shows firing rate density of each neuron, 
and red broken line represents average of firing rate density over all neurons in the ensemble. 
In the lower figures, binary activities are illustrated. 
One can observe very clear synchronous activities of neurons in the both ensembles. 
Therefore, we conclude that our proposed method can classify neurons with similar activities into the same ensemble.

These results suggest that our algorithm succeeds in reducing the dimension of large scale neuronal data. 
The activities of 1408 neurons are classified into 62 different patterns. 
In particular, focusing on neurons that form large size ensembles, 
82\% of all neurons are classified into one of 20 ensembles. 
This indicates that the majority of neural activities in the ALM of mice performing an object location discrimination task 
can be represented by only 20 patterns. Additionally, we can say our algorithm enables us to understand 
what information each neuronal ensemble represents in an object location discrimination task. 
The ensemble X is related to a memory of pole position, 
because neurons are active in Delay epoch and inactive in Response epoch. 
The ensemble Y is related to licking, because neurons in this ensemble are active in Response epoch.

\section{Discussion and perspective}
We proposed improved Bayesian inference algorithm for neuronal ensembles. 
To avoid inappropriate Bayesian inference result, 
we introduced the transition rule of multiple neurons to new ensemble and the idea of simulated annealing.
For simulated annealing, we introduced decay constant $\tau$ to control annealing schedules of transient probability 
in equation (\ref{qdecay}) and learning rate in equation (\ref{epsilonupdate}). By numerical analysis for synthetic data,
we found that blockwise neuronal ensemble structure can be obtained successfully by our method
even with small initial number of ensembles. 
We also compare the computational times between our proposed algorithm and the original, 
which indicate that our algorithm has advantage for inference of appropriate ensembles.

In this work, we focused only on inference of neuronal ensembles, and we did not study the detail of neural network 
structure like connection. However, we believe that our idea for ensemble inference will be helpful for understanding whole network structure including connection. By improvement of our method, Bayesian inference framework for further detail of network structure will be constructed.

Some issues are remained as future works. 
As mentioned in section \ref{Sec:ApplyReal}, real experimental data of neuronal activity is often continuous, 
not binary as in our formulation. In this work, for application of our algorithm we make binary data from 
estimated firing rate. However, such data processing may neglect partial information in neuronal activity. 
Natural idea to amend this point is to generalize our formalism to continuous activity data. For this purpose, 
we must consider how the generative model in this work should be modified.

\section*{Acknowledgments}
We appreciate comments from Giovanni Diana and Yuishi Iwasaki.
This work is supported by KAKENHI Nos. 18K11175, 19K12178, 20H05774, 20H05776 (KT), 20K06934 (KO),
and by the Collaborative Technical Development in Data-driven Brain Science, RIKEN CBS.
KO is also supported by the grant JP20dm0207001 (issued to Murayama lab at RIKEN).


\section*{References}
\bibliography{article}

\end{document}